\documentclass[sigconf, nonacm]{acmart}
\usepackage{xkeyval}
\AtBeginDocument{%
  \providecommand\BibTeX{{%
    Bib\TeX}}}

\usepackage{multirow}
\usepackage{float}
\usepackage{xspace}
\renewcommand{\shortauthors}{Trovato et al.}
\newcommand{\etal}{\emph{et al.}\ }
\newcommand{\tool}{\textsc{VisDocSketcher}}
\newcommand{\evaltool}{\textsc{AutoSketchEval}}

\def\BibTeX{{\rm B\kern-.05em{\sc i\kern-.025em b}\kern-.08em
    T\kern-.1667em\lower.7ex\hbox{E}\kern-.125emX}}
\usepackage[most]{tcolorbox}
\tcbset {
  base/.style={
    arc=0mm, 
    bottomtitle=0.5mm,
    boxrule=0mm,
    colbacktitle=black!10!white, 
    coltitle=black, 
    fonttitle=\bfseries, 
    left=2.5mm,
    leftrule=1mm,
    right=3.5mm,
    title={#1},
    toptitle=0.75mm, 
  }
}
\usepackage[framemethod=tikz]{mdframed}
\usepackage{enumitem}
\usepackage{pifont}

\definecolor{brandblue}{rgb}{0.34, 0.7, 1}
\newtcolorbox{mainbox}[1]{
  colframe=brandblue, 
  base={#1}
}

\newtcolorbox{subbox}[1]{
  colframe=black!30!white,
  base={#1}
}

\definecolor{antiquewhite}{rgb}{0.98, 0.92, 0.84}
\definecolor{arsenic}{rgb}{0.23, 0.27, 0.29}

\newmdenv[innerlinewidth=0.5pt, roundcorner=4pt,linecolor=arsenic,backgroundcolor=antiquewhite,innerleftmargin=6pt,
innerrightmargin=6pt,innertopmargin=6pt,innerbottommargin=6pt]{promptbox}

\begin{document}

\title{\tool: Towards Scalable Visual Documentation\\with Agentic Systems}

\author{Luis F. Gomes}
\affiliation{%
  \institution{Carnegie Mellon University}
  \city{Pittsburgh}
  \state{PA}
  \country{USA}
}
\email{lfgomes@andrew.cmu.edu}


\author{Xin Zhou}
\affiliation{%
  \institution{Singapore Management University}
  \country{Singapore}}
\email{xinzhou.2020@phdcs.smu.edu.sg}

\author{David Lo}
\affiliation{%
  \institution{Singapore Management University}
  \country{Singapore}}
\email{davidlo@smu.edu.sg}

\author{Rui Abreu}
\affiliation{%
 \institution{Faculty of Engineering, University of Porto}
 \city{Porto}
 \country{Portugal}}
 \email{rui@computer.org}


\begin{abstract}
  Visual documentation is an effective tool for reducing the cognitive barrier developers face when understanding unfamiliar code, enabling more intuitive comprehension. Compared to textual documentation, it provides a higher-level understanding of the system structure and data flow. Developers usually prefer visual representations over lengthy textual descriptions for large software systems. 
  Visual documentation is both difficult to produce and challenging to evaluate. Manually creating it is time-consuming, and currently, no existing approach can automatically generate high-level visual documentation directly from code. Its evaluation is often subjective, making it difficult to standardize and automate.
  To address these challenges, this paper presents the first exploration of using agentic LLM systems to automatically generate visual documentation. We introduce \tool, the first agent-based approach that combines static analysis with LLM agents to identify key elements in the code and produce corresponding visual representations.
  We propose a novel evaluation framework, \evaltool, for assessing the quality of generated visual documentation using code-level metrics.
  The experimental results show that our approach can valid visual documentation for 74.4\% of the samples. It shows an improvement of 26.7--39.8\% over a simple template-based baseline. Our evaluation framework can reliably distinguish high-quality (code-aligned) visual documentation from low-quality (non-aligned) ones, achieving an AUC exceeding 0.87. Our work lays the foundation for future research on automated visual documentation by introducing practical tools that not only generate valid visual representations but also reliably assess their quality. 
\end{abstract}



\maketitle

\section{Introduction}
Developing software is a collaborative process that relies on forming, sharing, and iterating on mental models of complex systems. Traditionally, developers have used informal sketches and diagrams as effective tools to reason about design, plan implementation, and communicate ideas with others~\cite{software_mangano_2010, go_cherubini_2007, visual_walny_2011, sketches_baltes_2017}.
These visual artifacts, often drawn on whiteboards,  played a key role not only in team-based discussions but also in individual workflows, helping developers align their common understanding and navigate evolving codebases more easily~\cite{maintaining_latoza_2006}.

Sketches are highly valuable during the \textit{planning and design} phases of the software development lifecycle, but easily lose relevance in the \textit{maintenance} phase~\cite{sketches_baltes_2017}. Even if an initial sketch is used to guide the implementation of the first version of the code, as the software evolves, the original visual representation quickly becomes outdated and disconnected from the actual system~\cite{roundtrip_baltes_2017}. This gap becomes especially challenging when new developers join the team and must reconstruct the underlying mental model in the absence of proper visualizations \cite{10.1145/3641822.3641873}.

Thus, it is critical to maintain up-to-date visual artifacts throughout the software lifecycle. These visual representations serve as a bridge between the evolving code and the mental models developers need to understand, navigate, and improve the system effectively.

\vspace{0.2cm}
\noindent\textbf{Motivation.} One of the constant challenges in Software Engineering is code maintenance. Developers spend significant time trying to understand existing code before making modifications or adding new features~\cite{6227188}. This challenge is amplified in the most recent development environments, where large portions of code are increasingly generated by AI tools \cite{10.1145/3664646.3676277}. Developers are shifting from being active authors of code to passive reviewers of generated solutions~\cite{10.1145/3709353}. However, this shift leads to a superficial review process: developers are often overconfident in inaccurate AI-generated code without fully understanding it~\cite{Klemmer_trust, 10.1145/3576915.3623157}.

While generating textual documentation is already a common use case for LLMs and an active area of research~\cite{yang2025docagent, luo_etal_2024_repoagent}, integrating visual documentation (i.e., sketches) of the generated code can lower developers' cognitive barrier to understand new code and unlock more intuitive, human-centric ways to convey code structure and behavior~\cite{10.1145/2858036.2858372, 5336433}.
Textual documentation often provides low-level explanations of specific code segments or functions~\cite{10.1145/3519312}. In contrast, visual representations can be used to represent a higher-level understanding of the system’s structure and data flow. Moreover, developers often prefer visuals over lengthy textual descriptions, as ``\textit{the more the developers deal with large systems, the more they feel a need for a tool that facilitates visualization}''~\cite{921708}.

Despite the usefulness of visual documentations, manually creating them is time-consuming, and their evaluation is usually subjective and hard to standardize and automate.
Moreover, to the best of our knowledge, no existing study has systematically explored how to develop automatic approaches for generating meaningful sketches from code and evaluating their quality automatically.
To fill this research gap, our work tackles key challenges in \textit{generating} visual documentation and \textit{evaluating} the generated artifacts automatically.

\vspace{0.2cm}
\noindent
\textbf{Our Work.} 
This paper explores the emerging role of the agentic systems in automatically generating diagrams that help developers understand code. We implement \tool, a code-to-sketch system that automatically produces informal, sketch-style diagrams from code. Our approach combines static analysis with Large Language Model (LLM) agents to identify key elements in the code and produce corresponding visual representations. In addition, we propose a new evaluation framework \evaltool, to systematically assess the quality of sketches when reference sketches are not present. We selected the Data Science domain for our evaluation due to its inherently visual nature, characterized by dataflows and visualizations, as well as the availability of related work and existing datasets that support our experimental setup.

This work aims to answer the following research questions:

\begin{itemize}[leftmargin=0pt]

\item [-] \textbf{RQ1}: 
What metrics and methods can approximate the evaluation of visual code documentation without relying on ground truth sketches?

\item [-] \textbf{RQ2}: How well does our automatic sketch generation system perform?

\item [-] \textbf{RQ3}: How is sketch generation quality affected when visualizing more complex software systems?

\end{itemize}

We evaluate \tool\textit{} on two Data Science Jupyter Notebook benchmarks with complementary characteristics: \textit{Visual Code Assistants Artifacts} \cite{gomes_2025_15118203} and \textit{DistillKaggle} \cite{distill_mostafavi}. \textit{Visual Code Assistants Artifacts} is a high-quality dataset derived from a user study involving 19 data scientists and includes ground-truth, human-authored diagrams. In contrast, while \textit{DistillKaggle} lacks ground-truth sketches, it features notebooks of varying complexity, making it suitable for evaluating the performance of sketch generators on more complex notebooks.

The experimental results show that our evaluation framework \evaltool \text{} demonstrates strong discriminative power in assessing the alignment between visual representations and their corresponding code. It reliably distinguishes aligned, human-made sketches from not-aligned alternatives, achieving an AUC above 0.87 and a Cliff’s Delta of 0.74. This indicates not only statistical significance but also a substantial effect size. These results support \evaltool as a robust proxy for evaluating how effectively a visualization reflects the underlying code.
Moreover, our multi-agent approach \tool\textit{} outperforms the baseline by 26.7--39.8\% in terms of the quality of generated sketches. We also find that the multi-agent variant outperforms the single-agent variant in 59.3\% of the cases. 
Lastly, the sketch generation quality declines significantly with notebook complexity: every 100 additional lines of code leads to an 8\% drop in CodeBLEU-Dataflow score. For every 10 Jupyter Notebook code cells, quality reduces by 9\%. These effects highlight the practical challenges of scaling sketch generation to complex notebooks.

\vspace{0.1cm}
\noindent
\textbf{Contributions.} 
The contributions of our work are as follows:
\begin{itemize}[left=2pt]
\item We introduce \tool, the first agentic system for generating \textit{informal}, high-level visual documentation \textit{directly} from source code. We empirically compare two design strategies for orchestrating the generation process: \textit{single-agent} versus \textit{multi-agent}.

\item We propose \evaltool, a novel evaluation framework to reduce the dependence on manually crafted ground-truth diagrams. It is inspired by self-supervised learning concepts and tested with human-made ground-truth diagrams.

\item Through our experimental analysis, we reveal key factors that influence the effectiveness of agent collaboration when generating visual documentation. We highlight the benefits and limitations of single and multi-agent coordination, suggesting practical guidance for designing agentic visual pipelines.

\item We release a comprehensive dataset of generated visuals, execution scripts, and intermediate representations to support reproducibility and further research.
\end{itemize}

\noindent \textbf{Data Availability.} All artifacts can be found at \url{https://anonymous.4open.science/r/VisDocSketcher-084F}.


\section{Related Work}

Understanding code is at the core of software development. Whether maintaining existing software systems or extending them with new functionality, developers must form accurate mental models of how a system works: e.g., how components interact, how data flows, and how behavior emerges from structure. These mental models are often externalized through visual artifacts, such as informal sketches or diagrams, which help developers plan, explain, and reason about code. 

\subsection{Visuals in SE and Code Assistants}

Previous work has investigated how developers use informal sketches to support software development. LaToza \etal~\cite{maintaining_latoza_2006} showed that developers construct complex mental models to understand code, emphasizing the need for tools that link informal sketches to code artifacts. Mangano \etal~\cite{software_mangano_2010} identified that developers use low-detail abstractions and mix notations when sketching. In the same line, Walny \etal~\cite{visual_walny_2011} traced how sketches evolve across tasks, showing that informal diagrams serve not only as design tools but also as memory aids and communication devices. Baltes \etal~\cite{sketches_baltes_2017} demonstrated that most sketches are informal, abstract, and created on analog media, proposing tools like SketchLink~\cite{linking_baltes_2017} and LivelySketches~\cite{roundtrip_baltes_2017} to bridge analog and digital workflows. Building on Mangano’s observations regarding the frequent use of mixed notations in developer sketches, Gomes \etal~\cite{gomes2025} recently explored the potential of LLMs as \textit{Visual Code Assistants}: tools that help express intent visually and generate corresponding code. Through a user study with ML practitioners, they demonstrated that LLMs can interpret informal whiteboard-style sketches and produce executable code, highlighting the feasibility of vision-assisted coding workflows.

While the prior work explores the capacity of LLMs to interpret and translate visual representations into code, it leaves another crucial aspect of Visual Code Assistants unexplored: their ability to explain code visually.


\subsection{Generative Visual Documentation}


\textbf{Visuals Generation.} When generating visual representations with LLMs, two main approaches can be distinguished. The first involves using multi-modal models to generate images directly at the pixel level, while the second relies on generating code in a visual language (e.g., Mermaid, DOT) that can be rendered into visuals afterward. \textit{Text-to-image} generation has seen significant advances in recent years, with methods evolving from Generative Adversarial Networks (GANs)~\cite{StackGAN, AttnGAN} to more stable and higher-quality approaches such as diffusion models~\cite{Imagen3} and transformer-based autoregressive models~\cite{DALLE}. All these models are very successful in generating photorealistic images. 
However, these models are not suitable for directly producing structured visual artifacts such as flowcharts or architecture sketches, which require precise spatial arrangements and semantically consistent representations~\cite{wei2025words}.

Similarly, \textit{text-to-code} generation has advanced rapidly, with both specialized models, such as StarCoder, CodeLLaMA,  DeepSeek-Coder, and general-purpose LLMs such as LLaMA3, Gemini, GPT-4o, or Claude, achieving strong performance on complex programming tasks and strong adaptability across languages. 
However, these models do not support direct sketch generation.

\vspace{0.1cm}
\noindent
\textbf{Multi-Agents.} A promising trend is the use of multiple specialized LLM agents working collaboratively, rather than relying on a single model~\cite{he2025llm}.

For instance, the multi-agent approach has led to improved performance in text visualization. Specifically, Wei \etal~\cite{wei2025words} presented \textit{DiagramAgent}, a text-to-diagram framework to generate and edit diagrams. 

Similarly, in the task of generating textual documentation for code, multi-agent systems have shown improved performance.
The most recent example is Meta’s DocAgent~\cite{DocAgent}, a code-to-text system consisting of a team of four specialized agents: reader, searcher, writer, and verifier.

The team is coordinated by an orchestrator, which delegates tasks to each of them. They also pointed out that traditional metrics require gold references, and while human evaluation is the most reliable method, it is costly, subjective, and difficult to scale. To tackle this problem, they proposed an evaluation framework using an LLM-as-judge with scoring rubrics, encompassing  \textit{Completeness}, \textit{Helpfulness}, and \textit{Truthfulness} of the generated documentation.

However, none of the existing approaches directly address the problem of automatically generating visual documentation from code. 

In this work, we propose \tool, the first agentic approach that combines static analysis with LLM agents to produce valid visual documentation.

\section{\tool: Automatic Visual Documentation Generator}

\subsection{Mermaid Diagram}
Mermaid diagrams are text-based visualizations generated from a lightweight markup language called Mermaid.js.\footnote{\url{https://mermaid.js.org}} Designed for simplicity, Mermaid enables the creation of structured diagrams, such as flowcharts or class diagrams, using plain-text descriptions. This format is particularly well-suited for automated generation and reproducible documentation, as it allows diagrams to be rendered programmatically without manual graphical editing. In this work, our approach first generates the corresponding Mermaid code, which is then rendered into visual documentation using the Mermaid rendering engine.

\begin{figure}
    \centering
    \includegraphics[width=1\linewidth]{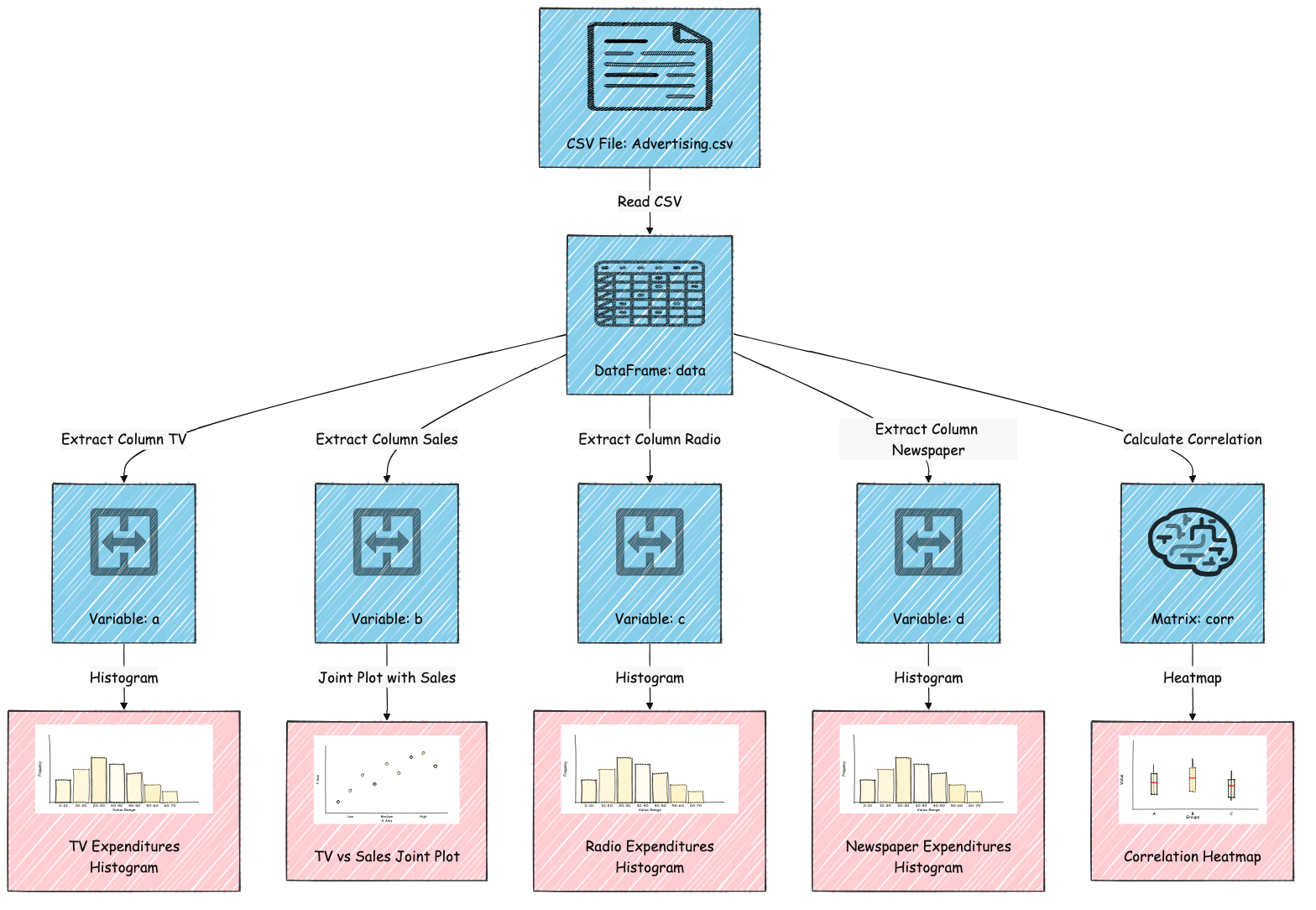}
    \caption{Example of a sketch generated by \tool.}
    \label{fig:mot_example}
\end{figure}

\begin{figure*}[ht]
    \centering
    \includegraphics[width=\linewidth]{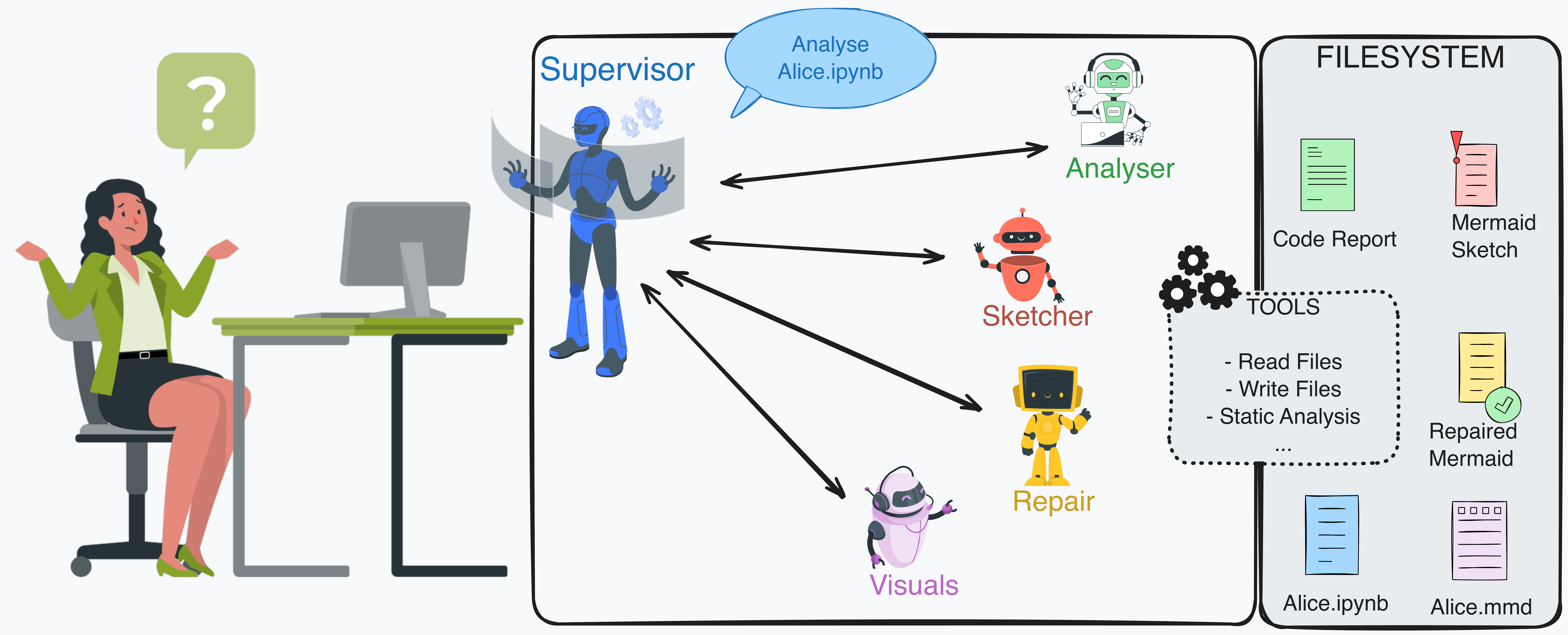}
    \caption{Overview of the multi-agent architecture. A supervisor agent receives the user query and orchestrates the workflow by delegating tasks to specialized agents. Each agent communicates exclusively with the supervisor (no inter-agent communication). Agents are equipped with tools to access and manipulate the local filesystem.}

    \label{fig:architecture}
    \Description{The architecture of the system.}
\end{figure*}
\subsection{Motivating Example}
In this section, we provide an example of the problem that was not explored in previous lines of research.

\vspace{0.1cm}
\noindent
\textbf{Scenario.} Alice has just joined a Data Science team that maintains a Jupyter-based Machine Learning workflow. Her first task is to understand a notebook that performs data pre-processing, model training, and evaluation across multiple cells. The notebook contains Python code interspersed with markdown and outputs, with several data transformations and model configurations scattered throughout. As a new intern, Alice faces some challenges:
\begin{itemize} [left=3pt]
    \item \textbf{Disjointed Logic}: The logic is split across multiple cells, making it hard to trace the flow of data.
    \item \textbf{No Overview:} There's no intuitive visual documentation that explains how the dataset moves from raw input to model evaluation.
    \item \textbf{Cognitive Load:} She has to mentally reconstruct the pipeline by reading and tracing code dependencies manually.
\end{itemize}

\vspace{0.1cm}
\noindent
\textbf{Proposed Solution.} In this scenario, a visualization of the steps of the notebook, including data sources, variable dependencies, and ML models' indications, can help Alice to understand the code. With a high-level visual summary, Alice can quickly grasp the big picture structure of the notebook, spot inconsistencies between the code and the expected flow, and ask informed questions before digging through every cell. In Figure \ref{fig:mot_example} we see an example of a diagram generated by \tool, which could be useful for Alice.

\vspace{0.1cm}
\noindent
\textbf{Challenges.} Developing this system involves several challenges. Firstly, it requires automatically extracting meaningful structural and semantic information from code to generate accurate and readable visualizations. Secondly, assessing the quality of these visualizations requires a robust evaluation framework that can identify whether the underlying high-level code structure is accurately represented by the diagram or not. 
In the following subsections, we explain in detail how we tackled each of these challenges.

\subsection{Single Agent Setup}

As the initial approach in \tool, we implemented a streamlined single-agent system responsible for converting Jupyter notebooks into high-level Mermaid diagrams. In our context, we use Mermaid flowcharts to represent the high-level workflow of Jupyter notebooks, capturing stages such as data loading, preprocessing, modeling, and evaluation.

In this setup, we extract the full content of each Jupyter notebook and provide it to the LLM agent embedded in a single prompt\footnote{Due to space constraints, we present a simplified prompt here. The full prompts, including those used for the multi-agent setup, are available in the replication package.}:

\begin{tcolorbox}[colback=gray!5, colframe=black!70, title=(Simplified) Prompt for Single Agent, width=\linewidth, boxrule=0.8pt, arc=2mm]
\textbf{[Role]} You are an expert software visualization system specialized in Python-based data science workflow...

\textbf{[Task Description]} Your task is to generate a high-level informal sketch of a Jupyter Notebook...

\textbf{[Output Format Control]}
The diagram should use the Mermaid flowchart syntax and focus on representing major steps such as data loading, preprocessing, visualization, modeling, and evaluation.
Use intuitive and include emojis in node labels...

\textbf{[Data]} Here is the notebook content: ...
\end{tcolorbox}

This agent does not have access to any tools; it receives the notebook content and returns a Mermaid diagram as a raw text response. The prompt instructs the model to generate a high-level flowchart of the notebook’s workflow, highlighting key stages such as data loading, preprocessing, modeling, and evaluation. The response is saved directly as a .mmd file (i.e., a plain text file written in Mermaid syntax that can be rendered into a Mermaid diagram).
The single-agent system also includes logic to handle missing files, skipped generations, and consistent output directory structures, enabling scalable and automated diagram generation across large notebook datasets.

\subsection{Multi Agent Setup}
Following successful work in other areas~\cite{yang2025docagent}, we built a second version of \tool \textit{} as a multi-agent system, where each agent is responsible for a specific task as shown in Figure~\ref{fig:architecture}. The architecture is implemented using LangGraph~\cite{LangGraph} relies on OpenAI's \textit{GPT-4o-mini} model to drive the reasoning capabilities of each agent\footnote{The choice of the underlying LLM is independent of our tool. For consistency and robustness in evaluation, we use GPT-4o-mini across all experiments.}. Each agent operates independently, maintaining its own reasoning prompt and toolset. The communication is always made through the supervisor, with intermediate artifacts being saved to the local filesystem.

\vspace{0.1cm}
\noindent
\textbf{Supervisor Agent.} At the core of the system is the supervisor agent, responsible for interfacing with the user and coordinating all other agents by directing when and how they should operate. It is the only agent that receives a user prompt; all remaining agents are guided exclusively through our system prompts, and by the supervisor prompts during execution. To monitor progress and ensure the correct completion of each step, the supervisor is equipped with a dedicated \texttt{file\_exists} tool, allowing it to verify the presence of generated artifacts throughout the workflow. All the other agents have two common tools: \texttt{write\_file} and \texttt{read\_file}.

\vspace{0.1cm}
\noindent
\textbf{Analyser Agent.} This agent processes Jupyter notebooks to extract key structural and semantic elements. It identifies data sources, logical sections of the notebook, and the flow of data variables across cells. In addition, it recognizes ML components such as model definitions, training routines, and evaluation steps. This agent uses a function to extract notebook cells and static analysis approaches, including the construction of data flow graphs and analysis of variable dependencies, to understand the notebook. Then it writes its analysis to an intermediate file for downstream agents. The tools that it can use are \texttt{build\_dataflow\_graph\_ast}, \texttt{extract\_variable\_dependencies} and \texttt{extract\_notebook\_cells}.

\vspace{0.1cm}
\noindent
\textbf{Sketcher Agent.} The sketcher consumes the analysis output from the analyser agent and generates an initial workflow diagram using Mermaid.js. This diagram focuses on the data flow and logical dependencies inferred from the code. The resulting sketch provides a clear, structural overview of the workflow.
It does not have extra tools.

\vspace{0.1cm}
\noindent
\textbf{Repair Agent.} Given that Mermaid.js syntax can be sensitive and sometimes error-prone, the repair agent is introduced to automatically identify and fix issues in the generated diagrams. This includes handling malformed syntax, escaping special characters, and ensuring that the final diagram is valid according to Mermaid specifications. The agent validates the syntax before saving the corrected version for further enhancement. It has access to \textit{validate\_mermaid\_from\_file} tool that allows the syntax verification of Mermaid.js files.

\vspace{0.1cm}
\noindent
\textbf{Visuals Agent.} To make the diagrams more intuitive and engaging, the repaired sketches are augmented with visual elements. This includes adding emojis to nodes, applying color schemes based on node types (e.g., data vs. model), and integrating dummy plots or icons that hint at the function of each step. This agent ensures visual consistency and improves readability without compromising the underlying structure. 



\vspace{0.1cm}
\noindent
\textbf{How Agents Collaborate.}
 In summary, when the user requests the generation of a new diagram from code, the Supervisor Agent first calls the Analyser Agent which processes the notebook; the Sketcher Agent produces an initial diagram based on this analysis, the Repair Agent corrects any syntactic issues, and the Visuals Agent enriches the output with visual elements; However the \textit{Supervisor} can decide to repeat any of the steps based on new information gathered from other agents (for instance, it can request the Repair Agent to act again after the Visuals).

 The modular structure of the system allows for easy extensibility, enabling researchers to add new agents or modify existing ones with minimal coupling. This ensures that similar systems can be adapted to other areas of SE, such as Software Architecture.

\subsection{Evaluation Framework}\label{sec:eval_framework}

\begin{figure*}[ht]
    \centering
    \includegraphics[width=\linewidth]{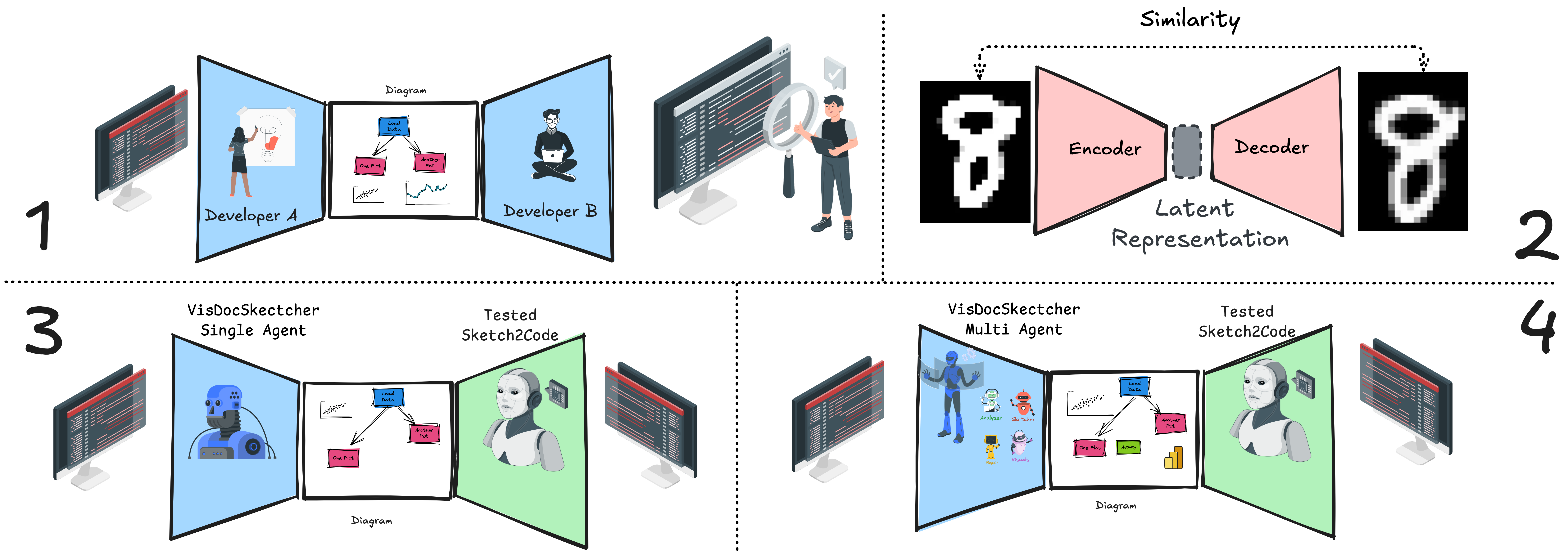}
    \caption{\textbf{Experimental Setup.}
    1) \textit{Human Reconstruction Analogy}: One developer generates a diagram from code, and another attempts to write code from it. If the code is similar, it means the diagram effectively captures the code's meaning. 
    2) \textit{Autoencoder Inspiration}: Autoencoders are models capable of reconstructing their input (e.g. images). High reconstruction similarity between input and output determines the quality of the model and it's components (Encoder and Decoder).
    3) \textit{Single-Agent Evaluation}: Fixing the decoder, the quality of the code reconstruction score depends solely on the capacity of the Code2Sketch Agent to generate sketches. Our single agent architecture is tested on its ability to generate meaningful sketches directly from code. 
    4) \textit{Multi-Agent Evaluation}: Using the same Sketch2Code \cite{gomes2025} model, we can compare the single and multi-agent architectures.}
    \label{fig:evaluation}
    \Description{Evaluation framework.}
\end{figure*}

One of the key challenges in evaluating the quality of generated visualizations is their inherent subjectivity: the same concept can be represented in multiple valid ways (e.g., loading a CSV file might be depicted as a box, a text label, or an icon). Additionally, there is a lack of labeled datasets that assess the quality of such sketches at the high-level of code flow and intent. 

\textbf{Inspiration.} As depicted in \ding{172} of Figure~\ref{fig:evaluation}, suppose we have two human developers: one who reads the code and produces a high-quality sketch (\textit{Code2Sketch}), and another who views only the sketch and is able to reconstruct the original code (\textit{Sketch2Code}). If the second developer successfully recovers the code, it is intuitive to conclude that the sketch is relevant and faithfully represents the original code. 
If we assume that we trust in the capabilities of the Sketch2Code developer, the whole reconstruction is dependent solely on the intermediate sketch and therefore on the capacity of the Code2Sketch developer to represent the code visually.
In Machine Learning, there is a similar concept, applying an unsupervised methodology that uses reconstruction with unlabeled datasets. \textit{Autoencoders}~\cite{li2023comprehensive} are neural networks designed to learn efficient data representations by compressing the input data into a lower-dimensional \textit{Latent Space} and then reconstructing it. As depicted in \ding{173} of Figure~\ref{fig:evaluation}, it is composed of two parts: the \textit{Encoder}, which reduces the dimensionality of the input data, creating a lower-dimensional representation, while the \textit{Decoder} reconstructs the original input from the compressed representation. The difference between the input data and the reconstructed data is the \textit{Reconstruction Similarity}~\cite{Autoencoders}. 

Translating the human scenario and inspired by the autoencoder concept, we employ a similar idea at the higher level of LLM agents. Our evaluation framework consists of two main components: a Code2Sketch mechanism, equivalent to the \textit{Encoder}, and a Sketch2Code tool, equivalent to the \textit{Decoder}. The counterpart of the \textit{Latent Space} is the generated visual documentation, which represents the code we want to reconstruct, containing less information than the original code. By choosing a tested Sketch2Code \textit{Decoder} from the literature, we can work with the following \textit{assumption: if the Sketch2Code tool receives a high-quality sketch, it will represent it with high fidelity in the reconstructed code}. Therefore, the final score of the framework depends entirely on the quality of the input sketch and, consequently, on the Encoder's ability to generate an accurate and meaningful sketch from the original code.

\section{Evaluation Design \label{sec:eval_design}}

\subsection{Research Questions}
Our work aims to answer three Research Questions (RQs). 
\begin{itemize}[leftmargin=*]

\item \textbf{RQ1: What metrics and methods can approximate the evaluation
of visual code documentation without relying on ground truth
sketches?}
In RQ1, we demonstrate the reliability of our visual documentation evaluation framework, \evaltool.

\item \textbf{RQ2: How well does our automatic sketch generation system perform?} 
In RQ2, we present the performance of our proposed agentic sketch generation approach \tool.

\item \textbf{RQ3: How is sketch generation quality affected when visualizing more complex software systems?}
In RQ3, we analyze how software complexity impacts the quality of sketches generated by \tool.
\end{itemize}

\subsection{Datasets \label{sec:datasets}}
Our framework is evaluated in the Data Science domain, which closely aligns with our use case and benefits from publicly available datasets that support comprehensive testing. We initially considered the DS-1000 dataset~\cite{ds_1000}, a benchmark containing one thousand data science problems. While its domain relevance and manageable size made it a promising candidate, a closer analysis revealed its limitations for our task: DS-1000 focuses on low-level code operations, whereas our framework targets high-level structural representations of code.

Instead, we used two recently published datasets with complementary characteristics: \textit{Visual Code Assistants Artifacts} \cite{gomes_2025_15118203} and \textit{DistillKaggle} \cite{distill_mostafavi}. \textit{Visual Code Assistants Artifacts} is derived from a user study involving 19 data scientists, and includes both \textit{ground-truth, human-authored diagrams} and the corresponding intended Jupyter notebooks. This makes it particularly valuable for the validation of our evaluation framework (RQ1) and to make comparisons between our agentic approaches with the baseline (RQ2). 

\textit{DistillKaggle} is a large-scale, curated dataset extracted from Kaggle notebooks, comprising over 500,000 files. It includes metadata such as \textit{Lines of Code}, \textit{Number of Visualizations}, and \textit{Developer Tier}. Although this dataset lacks ground-truth sketches, it is valuable for studying how sketch generation quality is influenced by notebook complexity and author expertise.
From the \textit{DistillKaggle} corpus, we cleaned and curated a sample of 1,000 Jupyter notebooks, balancing it across key metadata dimensions (Lines of Code, Number of Code Cells). This curated dataset is particularly appropriate for studying scalability effects (RQ3).

For simplicity, we refer to the \textit{Visual Code Assistants Artifacts} dataset as the \textbf{ground-truth dataset}, as it contains human-created sketches. Moreover, we refer to \textit{DistillKaggle} as the \textbf{large dataset}, since it includes more diverse Jupyter notebooks, although it does not contain ground-truth sketches.

\subsection{Baseline}
Since there are no established baselines for this task in the existing literature, we implemented a simple template-based baseline that extracts information from the notebook and generates a corresponding diagram aimed at representing the dataflow structure. Although this baseline captures some structural elements, it lacks the semantic richness of our approach. By comparing this baseline against our agent-based approaches, we demonstrate that meaningful sketch generation benefits significantly from code-aware reasoning and structure extraction.

\subsection{Metrics}

In this work, we adopt our proposed novel visualization evaluation framework \evaltool \text{} (introduced in Section~\ref{sec:eval_framework}) to assesss the generated sketches. As demonstrated later in Section~\ref{sec:RQ1}, our results will show that this evaluation framework is reliable. 

\evaltool \text{} uses the quality of the code reconstructed from a sketch as a proxy for evaluating the quality of the sketch itself. 
To assess the quality of reconstructed code, we follow prior work~\cite{ren2020codebleu,codebertscore} by comparing it to the original code.
Specifically, we adopt a combination of well-established and widely used code similarity metrics in the literature~\cite{zhou2025llmasjudgemetricbridginggap}: match-based and embedding-based metrics.

\vspace{0.1cm}
\noindent
\textbf{Match-based.}  
We employ CodeBLEU~\cite{ren2020codebleu}, a code-specific extension of the BLEU metric. CodeBLEU is composed of four sub-metrics: (1) \textit{n-gram match}, which measures token-level overlap; (2) \textit{weighted n-gram match}, which incorporates domain-specific weighting; (3) \textit{syntax match}, which compares syntactic structures using AST-based representations; and (4) \textit{data flow match}, which captures the semantic correctness of variable usage and dependencies. The final CodeBLEU score is a weighted combination of these components, offering a more nuanced similarity assessment than traditional BLEU. 
In this work, we use both \textit{CodeBLEU-full} (an average of all four components) and \textit{CodeBLEU-dataflow} (using only the data flow component), with the latter only reflecting high-level semantic similarity.

\vspace{0.1cm}
\noindent
\textbf{Embedding-based.}  
We use CodeBERTScore~\cite{codebertscore}, an embedding-based metric built on top of the CodeBERT model~\cite{codebert}. It computes similarity by embedding the original and reconstructed code snippets and comparing them in vector space. CodeBERTScore accounts for lexical and semantic similarity and includes precision, recall, F1, and F3 sub-metrics. This approach is particularly useful when the reconstructed code differs in surface form but maintains functional equivalence.
In this work, we use both \textit{CodeBERTScore-F1} and \textit{CodeBERTScore-Precision}, which are widely used in prior studies.

\subsection{Experimental Setup}

\vspace{0.1cm}
\noindent
\textbf{\evaltool \textit{} Framework Validation (Setup for RQ1).} 
A framework for determining the quality of subjective artifacts is valid if it is capable of mimicking a human evaluation of the same artifact. In our case, it should be able to distinguish what a human annotator considers to be a sketch \textit{aligned} or \textit{unaligned} with the code. Each visual documentation can \textit{correctly} or \textit{incorrectly} reflect the intentions of the developer and the corresponding code. Consequently, the framework is valid if it is able to separate these two cases accurately.

To evaluate our hypothesis, we pair each drawing from the \textit{ground-truth dataset} with two distinct sets of code notebooks. For the \textit{aligned} reconstruction condition, each drawing is matched to the corresponding solution notebook submitted by the same participant, ensuring a strong alignment between the diagram and the underlying implementation. For the \textit{unaligned} reconstruction condition, we deliberately pair each drawing with a randomly selected notebook, creating a mismatch between the visual representation and the code. Significance and magnitude tests, such as AUC \cite{auc_test}, are performed to verify the separation capability of the framework. 

\vspace{0.1cm}
\noindent
\textbf{\tool \textit{} Evaluation (Setup for RQ2).} 

We use the high-quality \textit{ground-truth dataset} to compare our proposed tool, \tool, against the baseline. Specifically, both \tool \text{} and the baseline generate sketches for each Jupyter Notebook, and the quality of these sketches is assessed using our evaluation framework, \evaltool, whose reliability is established in the results of RQ1.

\vspace{0.1cm}
\noindent
\textbf{Impact of Input Complexity (Setup for RQ3).} To investigate how sketch generation quality is influenced by notebook complexity, we conduct experiments on the \textit{large dataset}, which contains notebooks with varying lengths and structural complexity. This dataset is well-suited for analyzing the impact of input complexity.
Specifically, we conduct a linear regression analysis using Lines of Code (LOC) and Number of Code Cells (CC) as independent variables, while controlling for developer expertise through the Performance Tier (PT) variable. As the dependent variable we select the \textit{Dataflow} metric as it captures dependency relations among variables \cite{ren2020codebleu}, which is the most relevant to what we want to see when building a representation of a data science workflow, while demonstrating strong discriminative power (see Table~\ref{tab:baseline}).

\section{Results}

\subsection{RQ1: Validating the Evaluation Framework}\label{sec:RQ1}

In RQ1, we validate our evaluation framework, \evaltool, by assessing its reliability in distinguishing sketches that are aligned with the code from those that are not. As introduced in Section 4.5, we curated a dataset from the ground-truth data specifically for this purpose, containing both aligned sketches paired with their corresponding code and unaligned sketches with unrelated code.

\begin{figure}[ht]
    \centering
    \includegraphics[width=0.9\linewidth]{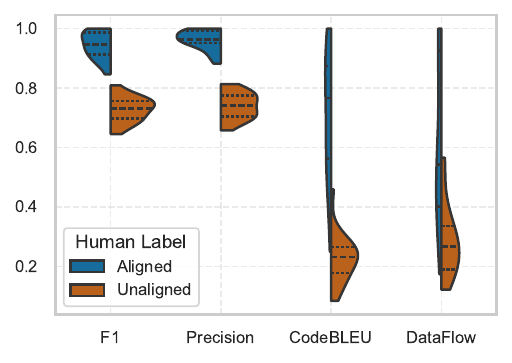}
    \caption{\evaltool \textit{} Framework Validation Results.}
    \label{fig:validation_framework}
    \Description{Framework Validation Results.}
\end{figure}

\vspace{0.1cm}
\noindent\textbf{Results.}
Figure~\ref{fig:validation_framework} presents the distribution of evaluation scores measured by CodeBERTScore-F1, CodeBERTScore-Precision, CodeBLEU-full, and CodeBLEU-Dataflow, for code reconstruction in two scenarios: sketches well aligned with the reference code and randomly sampled sketches lacking such alignment. The result confirms that our proposed evaluation framework can effectively distinguish aligned sketches from unaligned ones using code metrics (e.g., CodeBLEU) of the code reconstructed from sketches. However, we can also identify that different metrics present distinct abilities to discriminate between the two groups. Specifically,  some distributions present more overlap than others.

\begin{table}[ht]
\centering
\caption{\evaltool \textit{} – Statistical Results for Metric-Based Differentiation Between Aligned and Unaligned Samples.}
\label{tab:evaluation_results}
\footnotesize
\begin{tabular}{lllll}
\toprule
\textbf{Source} & \textbf{Metric} & \textbf{↓ Levene\_p} & \textbf{↑ AUC} & \textbf{↑ Cliff's $\Delta$} \\
\midrule
CodeBERTScore & F1            & 0.672    & 1.000 & 1.00  \\
CodeBERTScore & Precision     & 0.113    & 1.000 & 1.00  \\
\textbf{CodeBLEU} & \textbf{CodeBLEU}      & \textbf{0.000***} & \textbf{0.978} & \textbf{0.96}  \\
\textbf{CodeBLEU} & \textbf{Dataflow}      &\textbf{0.003***} & \textbf{0.871} & \textbf{0.74}  \\
\bottomrule
\end{tabular}
\end{table}

To evaluate which of the selected metrics more significantly distinguishes the two cases, we applied three complementary non-parametric statistical tests: the \textit{Mann-Whitney U} test~\cite{mann_test}, the \textit{Levene’s test}~\cite{levene_test}, and the \textit{Kolmogorov–Smirnov (KS)} test~\cite{ks_test}. 
The test results are summarized in Table~\ref{tab:evaluation_results}. Both the Mann–Whitney U and Kolmogorov–Smirnov tests show statistical significance for all metrics ($p$-value~$<$~0.001); therefore, their detailed results are omitted from the table. Additionally, Levene’s test was significant for CodeBLEU and CodeBLEU-Dataflow, indicating that the variance in scores between aligned and unaligned sketches differs significantly for these metrics.
Based on these results, CodeBLEU and CodeBLEU-Dataflow emerge as the most suitable metrics, as their ability to distinguish between aligned and unaligned sketches is supported by all three statistical tests.

It is also important to understand the magnitude of the separation between the two cases. The bigger the separation in this binary setup, the better. The Area Under the Curve (AUC)~\cite{auc_test} measures how well a metric can distinguish between two classes, in this case, aligned vs. unaligned. An AUC of 1.0 means perfect separation, while 0.5 indicates no better than random chance, given the equal number of good and bad cases in this experiment. Cliff’s Delta quantifies the effect size by measuring how often values in one group are higher than in the other. It ranges from -1 to 1, where 0 means no difference, and values above 0.47 are considered large effects. Together, AUC and Cliff’s Delta show not just whether the groups differ, but how strongly and consistently they do. The results of group separation are also presented in Table~\ref{tab:evaluation_results}. The results show that the evaluation framework \evaltool \text{} achieves consistently high AUC values ($\ge$ 0.87) and Cliff's $\Delta \ge 0.74$, representing a strong ability to discriminate between different classes.

\begin{mainbox}{Finding 1: Reliable Sketch Evaluation Framework}
Our proposed evaluation framework reveals a strong separation between high and low quality sketches, achieving consistently high \textbf{AUC values ($\ge$ 0.87) and Cliff's $\Delta \ge$ 0.74.}
\textbf{Sketch quality can be reliably assessed using only code-level metrics, without requiring ground-truth diagrams.} It possible to scale evaluation and flag low-quality sketches in practical cases.
\end{mainbox}

\subsection{RQ2: Performance of Automatic Sketch Generators}\label{sec:RQ2}

In RQ2, we present the performance of our proposed agentic sketch generation approach \tool \textit{}, and compare it with the baseline introduced in Section~4.3. In RQ2, we use the high-quality \textit{ground-truth dataset} as the evaluation data, and use the evaluation framework \evaltool \textit{} as the evaluation approach.

\begin{table}[t]
\centering
\caption{Comparison of baseline and \tool \textit{} performance, reporting the 5th percentile and average values.}
\label{tab:baseline}
\begin{tabular}{lllll}
\toprule
Metric    & \multicolumn{2}{c}{Perc. 05}                 & \multicolumn{2}{c}{Average}                \\
          & Baseline          & \texttt{VisDocS}             & Baseline          & \texttt{VisDocS}            \\
\midrule
F1        & 0.633             & \textbf{0.753}             & 0.690             & \textbf{0.836}             \\
precision & 0.725             & \textbf{0.784}             & 0.774             & \textbf{0.863}             \\
CodeBLEU  & 0.067             & \textbf{0.177}             & 0.132             & \textbf{0.333}             \\
dataflow  & 0.026             & \textbf{0.125}             & 0.103             & \textbf{0.342 }            \\
Average   & 0.363             & \textbf{0.460}             & 0.425             & \textbf{0.594}             \\
\bottomrule
\end{tabular}
\vspace{-0.4cm}
\end{table}

\vspace{0.2cm}\noindent\textbf{Results.} Table~\ref{tab:baseline} reports the mean and 5th percentile scores across several metrics, comparing the baseline to the average performance of \tool \textit{} (averaging single-agent and multi-agent variants). The 5th percentile indicates a conservative performance bound: 95\% of outputs achieve this score or higher, offering insight into worst-case reliability. Across all metrics, \tool \textit{} significantly outperforms the baseline, with all p-values below $10^{-6}$. On average, \tool \textit{} achieves a 39.8\% relative improvement in mean score and a 26.7\% relative improvement in the 5th percentile, demonstrating not only higher average quality but also more consistent and robust output. Even the weakest outputs from \tool \textit{} exceed the performance of most baseline results, indicating a substantial reduction in failure cases and more dependable visual documentation quality.

\begin{mainbox}{Finding 2: Improvement Over Baseline}
\tool \textit{} significantly outperforms the template-based baseline across all evaluated metrics. On average, \textbf{\tool \textit{} improves the mean semantic accuracy by 39.8\% and the 5th percentile by 26.7\%.} These gains highlight both higher overall quality and greater reliability, as 95\% of \tool's outputs perform above this robust lower bound.
\end{mainbox}

\begin{figure}[t]
    \centering
    \includegraphics[width=\linewidth]{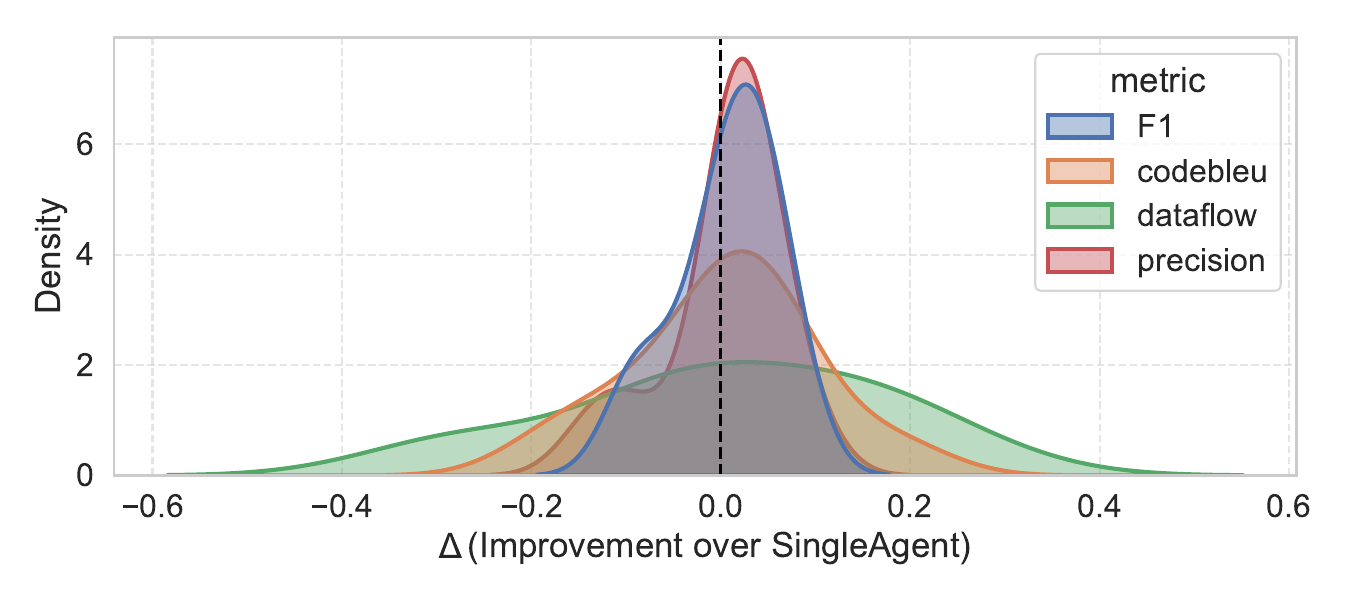}
    \caption{Distribution of Pairwise Differences Across Metrics. All metrics show a slight rightward shift ($>0$), reflecting consistent but modest improvements of Multi-Agent over Single-Agent architectures.}
    \label{fig:distribution_improvement}
    \vspace{-0.5cm}
\end{figure}

\noindent\textbf{Single Agent v.s. Multi-Agent.} 
We also compare the performance of the two \tool variants: the single-agent and multi-agent approaches.
Specifically, we compute pairwise differences for each metric $i$ with $\Delta_i = \text{score}_{\text{MA}}^i - \text{score}_{\text{SA}}^i$. As shown in Figure~\ref{fig:distribution_improvement}, the Multi-Agent system significantly outperforms the Single-Agent in \textbf{59.3\% of cases (p-value = 0.0335)}.

However, there is a significant disparity in processing time between the two architectures. On average, the single-agent system completes generation in 8.27 seconds (SD = 7.84), whereas the multi-agent setup takes 93.43 seconds (SD = 29.24), making it 11.3 times slower. This substantial difference reflects the coordination overhead inherent in multi-agent workflows, where specialized agents communicate and validate intermediate outputs. While this design yields higher-quality visual documentation, it introduces a notable computational cost that must be considered when deploying at scale.

\begin{mainbox}{Finding 3: Performance and Computation Trade-off}
The \textbf{multi-agent system outperforms the single-agent system in 59.3\% of the cases}. This improvement, however, comes at a substantial cost: the multi-agent setup is approximately \textbf{11.3 times slower}. These results highlight a fundamental trade-off between visual quality and system efficiency.
\end{mainbox}

\subsection{RQ3: Impact of Notebook Complexity}
As notebooks grow in size and structure, the task of generating accurate and complete visual documentation becomes increasingly challenging. In RQ3, we show how the complexity of a notebook affects \tool's ability to generate high-quality sketches, focusing on two factors: notebook size in \textit{Number of Lines of Code} (LOC) and notebook segmentation measured as the \textit{Number of Code Cells} (CC). We control the experiments for the developers' Kaggle Performance Tier (PT) as a proxy for their expertise.

To better understand how notebook complexity affects sketch generation quality, we first built a general regression model that estimates the main effects of complexity features. We then extended this model with interaction terms to assess whether these effects differ between the single-agent and multi-agent architectures.

Although CodeBLEU scores are often low in absolute terms due to their strict enforcement of structural similarity, our analysis focuses on the relative differences across complexity levels and model variants. We use the \textit{Dataflow} component of CodeBLEU as our independent variable, since it best captures the semantic structure of a notebook by reflecting how variables and operations are connected. This aligns with our goal of representing the code’s high-level workflow. The \textit{CodeBLEU-Dataflow} score shows a strong negative correlation with notebook complexity, -0.46 with lines of code (LOC), and -0.24 with number of code cells (CC), being a first indicator that as notebooks grow more complex, maintaining accurate relationships in the sketches becomes increasingly difficult.

\begin{table}[b]
\centering
\caption{Regression Results: Dataflow and Notebook Complexity.}
\label{tab:regression_summary}
\begin{tabular}{lrrrrrr}
\toprule
\textbf{Variable} & \multicolumn{3}{c}{\textbf{General Model}} & \multicolumn{3}{c}{\textbf{Interaction Model}} \\
\cmidrule(lr){2-4} \cmidrule(lr){5-7}
 & \textbf{Coef.} & \textbf{Std. Err.} & \textbf{\textit{p}} & \textbf{Coef.} & \textbf{Std. Err.} & \textbf{\textit{p}} \\
\midrule
Intercept                &  0.2654 & 0.0083 & ***  &  0.2688 & 0.0107 & ***  \\
multi                   &    ---  &   ---  & ---  & -0.0074 & 0.0168 &      \\
PT=1                   & -0.0074 & 0.0094 &      &  0.0041 & 0.0123 &      \\
PT=2                   & -0.0167 & 0.0085 &      &  0.0007 & 0.0110 &      \\
PT=3                   & -0.0146 & 0.0104 &      & -0.0029 & 0.0135 &      \\
PT=4                   &  0.0161 & 0.0125 &      &  0.0128 & 0.0162 &      \\
PT=5                   & -0.1500 & 0.0103 & ***  & -0.1550 & 0.0132 & ***  \\
PT=1 $\times$ multi    &    ---  &   ---  & ---  & -0.0268 & 0.0188 &      \\
PT=2 $\times$ multi    &    ---  &   ---  & ---  & -0.0437 & 0.0171 & *    \\
PT=3 $\times$ multi    &    ---  &   ---  & ---  & -0.0279 & 0.0208 &      \\
PT=4 $\times$ multi    &    ---  &   ---  & ---  &  0.0080 & 0.0252 &      \\
PT=5 $\times$ multi    &    ---  &   ---  & ---  &  0.0124 & 0.0209 &      \\
LOC                     & -0.0008 & 0.0000 & ***  & -0.0009 & 0.0001 & ***  \\
LOC $\times$ multi      &    ---  &   ---  & ---  &  0.0001 & 0.0001 &      \\
CC                      & -0.0012 & 0.0003 & **   & -0.0010 & 0.0004 & **   \\
CC $\times$ multi       &    ---  &   ---  & ---  & -0.0003 & 0.0006 &      \\
\midrule
\textbf{$R^2$}          & \multicolumn{3}{c}{0.3535} & \multicolumn{3}{c}{0.3697} \\
\textbf{Adjusted $R^2$} & \multicolumn{3}{c}{0.3504} & \multicolumn{3}{c}{0.3631} \\
\bottomrule
\end{tabular}
\small
\textit{Note}: $*$ $p < .05$, $**$ $p < .01$, $***$ $p < .001$
\end{table}

As shown in Table~\ref{tab:regression_summary}, both models explain a similar proportion of variance in the CodeBLEU-Dataflow metric (0.35 - 0.37). Given the complexity of factors influencing sketch quality and the limited set of predictors, explaining over one-third of the variance is a meaningful result. 

\textbf{Results.} In both models, the number of lines of code (LOC) and the number of code cells (CC) show statistically significant negative associations with the \textit{CodeBLEU-Dataflow} score. These effects are observed in both architectures, with ($p < 0.001$ for LOC; $p < 0.01$ for CC) providing strong evidence of their influence. The performance Tier (PT) also exhibits a significant effect, particularly for the highest tier (PT=5), which shows a strong negative coefficient, suggesting that the more experienced developers create more difficult-to-interpret code. Code authored by these experienced developers is harder to interpret visually, resulting in a \textbf{15.5 percentage point decrease} in dataflow quality compared to lower tiers.

\begin{mainbox}{Finding 4: Complexity Impacts Visual Quality}
Generation quality declines with increasing notebook complexity. \textbf{Every additional 100 lines of code leads to a 0.08 reduction in CodeBLEU-Dataflow, and every 10 additional code cells result in a 0.09 drop}. More experienced developers produce code that is significantly harder to visualize, with a 15.5 percentage point lower DataFlow score.
\end{mainbox}

While the other absolute coefficient values may initially appear small, they can be misleading without practical context. For example, an increase of \textbf{100 lines of code corresponds to an 8\% drop} in the dataflow score, and an increase of just \textbf{10 code cells leads to a 9\% reduction}. These are substantial effects in realistic scenarios, where it is natural to exceed these thresholds, reinforcing the practical relevance of LOC and CC as complexity indicators.

One contributing factor is that longer prompts can degrade the reasoning capabilities of LLMs. To investigate whether separating concerns can help address this issue, we examine the interaction effects. The interaction terms \texttt{LOC$\times$multi} and \texttt{CC$\times$multi} are not statistically significant, suggesting that the influence of notebook size and complexity on dataflow quality is consistent across both single-agent and multi-agent conditions. This indicates that the multi-agent setup neither amplifies nor mitigates the degradation in sketch quality caused by increased complexity.

\begin{mainbox}{Finding 5: Equal Sensitivity Across Architectures}
Interaction terms between system type and complexity are not statistically significant (p-value > 0.1). \textbf{Notebook complexity has a similar negative effect on visual documentation quality for both single-agent and multi-agent systems.}
\end{mainbox}

\noindent\textbf{Rendering Statistics.} 
We also examined rendering accuracy on the large dataset, as its increased complexity may prevent agentic models from generating valid Mermaid files, leading to rendering errors.
We found that the single-agent system successfully generated 999 Mermaid files, of which 849 were rendered without compilation errors (84.9\%). Unlike a fixed linear pipeline, the multi-agent system operates through flexible stages coordinated by the supervisor agent. This agent decides which intermediate steps to include based on the complexity of the code. For instance, it may skip or repeat some of the stages. While this flexibility enables adaptive behavior, it also introduces variability in the outputs due to both agent failures and supervisor decisions. In the large dataset, the multi-agent system generated 996 JSON reports and 990 initial sketches. Of these, 87.1\% (863) were repaired. 920 were further processed to include visual elements. Ultimately, in the multi-agent case, \textbf{74.4\%} of the sketches were rendered successfully without errors, producing valid visual representations.
Although the multi-agent variant generates fewer sketch predictions due to its collaborative nature, its sketches are of higher quality than those produced by the single-agent variant, as shown in the results of RQ2.

\section{Discussion}

\noindent
\textbf{Prompts for Visual Diagrams.} Although this work does not directly study the impact of visual components on human perception, we recognize their importance from a human-centric standpoint. Visuals are inherently subjective but can play a critical role in supporting comprehension. While adding visuals introduces additional complexity, such as increased runtime, longer prompts, and potential noise that may reduce performance on our results based on code-centric metrics, we consider it a valuable trade-off. To support future work on human-in-the-loop or usability-focused evaluations, we formulated the prompts to encourage visually appealing diagrams. This includes explicit instructions for the use of icons, images, and emojis, as well as color cues to differentiate node types (e.g., data vs. plot elements).

\vspace{0.2cm}
\noindent
\textbf{Implications for Code Assistant Development.}
Integrating an automatic evaluation mechanism such as \evaltool \textit{} can serve dual purposes: (1) guiding the assistant itself to iteratively improve the generated visualizations, and (2) providing developers with a confidence score in the UI that reflects how faithfully the sketch represents the underlying code. This has several practical implications:

\vspace{0.2cm}
\noindent
\textbf{Implications for Researchers.}
The proposed evaluation framework opens new directions for studying AI-generated visualizations across domains such as software architecture, data pipelines, and educational tooling. It enables systematic analysis of sketch fidelity and lays the groundwork for cross-domain adaptation, since allows evaluation framework can be extended to other structured artifacts (e.g., UML, database schemas). Researchers can also explore collaborative settings where structured reconstruction metrics help humans intervene effectively during diagram refinement.

\section{Threats to Validity}

\textbf{External Validity.}
Our study focuses on data science code. While being an important part of Software Engineering \cite{10.1145/3180155.3182515,10.1145/2884781.2884783}, it may not generalize to other software domains. The structure and abstraction of Jupyter notebooks differ significantly from full-scale software systems, where diagrams often reflect complex architectural layers, components, and long-term evolution. For example, generating a dataflow diagram from a single notebook is fundamentally different from producing architectural views of a distributed system. While we expect our framework to be adaptable, future work should validate its effectiveness in broader and more diverse software engineering contexts.

\vspace{0.2cm}
\noindent
\textbf{Internal Validity.}
One potential threat to internal validity is the absence of large-scale user studies to directly validate our assumptions about sketch quality and reconstruction fidelity. To mitigate this, we grounded our evaluation using human-authored sketches that align with their corresponding notebooks, ensuring a consistent and meaningful target for the reconstruction task. Further empirical validation involving developers in real-world or controlled experiments strengthen confidence in our findings.

\vspace{0.2cm}
\noindent
\textbf{Construct Validity.}
Threats to construct validity arise from how well our chosen evaluation metrics capture the quality and usefulness of a generated sketch. Although we adopt a dataflow-aware variant of CodeBLEU, which emphasizes structural and semantic alignment with the original code, it may not fully reflect all aspects that developers find valuable in a visual explanation (e.g., intuitiveness, layout clarity, or aesthetic presentation). To mitigate this, we conducted a manual inspection of the generated diagrams in the ground-truth dataset. Our observations support the results: baseline sketches tend to be overly simplistic and lack relevant code semantics, whereas \tool \textit{} consistently produces diagrams that align more closely with the underlying code and convey meaningful structure.

\section{Conclusion and Future Work}
We introduced a novel LLM agentic framework for generating high-level visual documentation of code, with a focus on dataflow reconstructions in the data science domain. Our agentic approach decomposes the generation task across specialized agents, coordinated by a central supervisor, enabling modular and interpretable sketch construction.
To address the lack of evaluation standards in assessing sketches, we proposed a novel, automatic evaluation framework grounded in established code-to-sketch similarity metrics. Using this framework, we conducted a comparative analysis between our approach and a template-based baseline. Our approach improves the baseline results by 26.7\% in the mean of our metrics (CodeBLEU and CodeBertScore).
The multi-agent system outperformed the single-agent system in 59.3\% of cases, but at the cost of higher processing time ($11.3\times$ slower on average).
Quantitative results indicate that sketch quality degrades as code complexity increases, with significant correlations between quality scores and structural notebook features such as lines of code and number of cells. This highlights both the limitations of current generation techniques and the importance of robust evaluation methods.
Our work opens promising directions for future research in visual generation for software engineering, including improving scalability, enabling richer forms of sketches, and extending to other domains such as software architecture or scientific workflows. 

In future work, we are interested in investigating the extent to which the generated sketches can assist with real-world onboarding scenarios.

\bibliographystyle{ACM-Reference-Format}
\bibliography{refs}

\end{document}